\documentclass[11pt]{article}

\usepackage[bitstream-charter,greekuppercase=upright,cal=cmcal]{mathdesign}
\usepackage[T1]{fontenc}

\usepackage{amsmath,amsthm,mathabx}

\usepackage{units}

\usepackage[affil-it]{authblk}
\usepackage[english]{babel}

\usepackage[margin=2.2cm]{geometry}
\usepackage{url}
\usepackage{verbatim}
\usepackage{xcolor}
\definecolor{dullmagenta}{rgb}{0.5,0,0.4}
\definecolor{dullblue}{rgb}{0.0,0,0.86}
\usepackage{hyperref}
\hypersetup{colorlinks=true,citecolor=dullblue,linkcolor=dullblue,filecolor=dullblue,urlcolor=dullblue,breaklinks=true,bookmarksnumbered=true, bookmarksopen=true,bookmarksopenlevel=1}
\allowdisplaybreaks
\usepackage{braket}

\usepackage{titlesec}

\titleformat*{\section}{\large \bfseries}
\titleformat*{\subsection}{\normalsize\bfseries}

\titlespacing\section{0pt}{12pt plus 4pt minus 2pt}{2pt plus 2pt minus 2pt}

\DeclareFontFamily{U}{bbold}{}
\DeclareFontShape{U}{bbold}{m}{n}
 {
  <-5.5> s*[1.069] bbold5
  <5.5-6.5> s*[1.069] bbold6
  <6.5-7.5> s*[1.069] bbold7
  <7.5-8.5> s*[1.069] bbold8
  <8.5-9.5> s*[1.069] bbold9
  <9.5-11> s*[1.069] bbold12 
  <11-15> s*[1.069] bbold12
  <15-> s*[1.069] bbold17
 }{}

\DeclareRobustCommand{\id}{%
  \text{\usefont{U}{bbold}{m}{n}1}%
}

\usepackage[backend=biber,sorting=none,style=phys-jmr,biblabel=brackets,backref=true,bibencoding=utf8,maxnames=20,eprint=true]{biblatex}
\makeatletter
\pretocmd{\blx@head@bibintoc}{\phantomsection}{}{\ddt}
\makeatother
\DefineBibliographyStrings{english}{%
  backrefpage = {page},
  backrefpages = {pages},
}
\usepackage{csquotes}

\DeclareMathOperator{\tr}{Tr} 
\usepackage{braket}
\newcommand{\ketbra}[1]{\ket{#1}\bra{#1}}

\usepackage{titling}
\pretitle{\begin{center} \large \bfseries}
\posttitle{\par\end{center}}
\predate{}
\postdate{}
\preauthor{\begin{center} \normalsize}
\postauthor{\end{center}}
\begin{document}

\title{Achievable error exponents of data compression with quantum side information\\ and communication over symmetric classical-quantum channels}

\author{Joseph M.~Renes}
\affil{\small Institute for Theoretical Physics, ETH Z\"urich, 8093 Z\"urich, Switzerland}

\date{}
\maketitle

\renewcommand{\abstractname}{\vspace{-1.25\baselineskip}} 
\begin{abstract}
A fundamental quantity of interest in Shannon theory, classical or quantum, is the optimal error exponent of a given channel $W$ and rate $R$: the constant $E(W,R)$ which governs the exponential decay of decoding error when using ever larger codes of fixed rate $R$ to communicate over ever more (memoryless) instances of a given channel $W$. Here I show that a bound by Hayashi [\href{https://doi.org/10.1007/s00220-014-2174-y}{CMP 333, 335 (2015)}] for an analogous quantity in privacy amplification  implies a lower bound on the error exponent of communication over symmetric classical-quantum channels. 
The resulting bound matches Dalai's [\href{https://doi.org/10.1109/TIT.2013.2283794}{IEEE TIT 59, 8027 (2013)}] sphere-packing upper bound for rates above a critical value, and reproduces the well-known classical result for symmetric channels. 
The argument proceeds by first relating the error exponent of privacy amplification to that of compression of classical information with quantum side information, which gives a lower bound that matches the sphere-packing upper bound of Cheng et al.\ [\href{https://doi.org/10.1109/TIT.2020.3038517}{IEEE TIT 67, 902 (2021)}]. 
In turn, the polynomial prefactors to the sphere-packing bound found by Cheng et al.\ may be translated to the privacy amplification problem, sharpening a recent result by Li, Yao, and Hayashi [\href{http://arxiv.org/abs/2111.01075}{arXiv:2111.01075 [quant-ph]}], at least for linear randomness extractors. 
\end{abstract}
\vspace{1mm}

\section{Introduction}
When communicating over a classical channel $W$ at a rate $R$ below the capacity, a good code will have a decoding error probability which decays exponentially in the blocklength of the code. 
The optimal decay is characterized by the error exponent, the largest $E(W,R)$ such that the probability of error scales as $2^{-n\,E(W,R)}$ for blocklengths $n\to \infty$. 
This quantity is also known as the reliability function. 
Nearly matching lower and upper bounds on the error exponent for classical channels were first established by the lower bounds of Fano \cite{fano_transmission_1961} and Gallager~\cite{gallager_simple_1965} and the {sphere-packing} upper bound of Shannon, Gallager, and Berlekamp~\cite{shannon_lower_1967}. 
The bounds match precisely for rates above a certain critical value. 

Less is known about error exponents of channels with a classical input but quantum output (CQ channels).
Burnashev and Holevo initiated the study of the error exponent and found a lower bound on the error exponent for the case that the channel has pure state outputs~\cite{burnashev_reliability_1998,holevo_reliability_2000}. Further bounds were given by Hayashi~\cite{hayashi_error_2007,hayashi_universal_2009} and Dalai~\cite{dalai_note_2017}. 
Upper bounds on the error exponent were given by Winter \cite{winter_coding_1999-1} and Dalai~\cite{dalai_lower_2013}, with follow-up work by them in \cite{dalai_constant_2017,dalai_remarks_2017}. Further refinements to the bound for finite-blocklength were made by Cheng, Hsieh, and Tomamichel~\cite{cheng_quantum_2019}.
Notably, the sphere-packing bound of Dalai follows \cite{shannon_lower_1967} and shows that the setting of CQ channels also encompasses Lov\'asz' bound on the zero-error capacity of a channel~\cite{lovasz_shannon_1979}. 

The reliability of error correction has long been known to be related to the secrecy of privacy amplification in the quantum setting. 
This relation is exploited in security proofs of quantum key distribution in~\cite{lo_unconditional_1999,shor_simple_2000}, for instance. 
In \cite{renes_duality_2018} I showed a precise relation between the average decoding error probability when using a linear code for communication over a CQ channel and a particular security parameter of an associated linear extraction function employed for privacy amplification of a certain ``dual'' CQ state. 
The security parameter is measured in terms of the fidelity (or equivalently, purified distance) to the nearest ideal output. 
This indicates that the error exponents of the two tasks are identical, at least when employing linear codes or linear extractors, respectively. 
As in the case of channel coding, there exist privacy amplification protocols such that the security parameter decays exponentially; Hayashi provides a lower bound on the exponent in~\cite{hayashi_precise_2015}. 
Recently, a nearly matching upper bound on the exponent was established in~\cite{li_tight_2022-1}, which agrees with that lower bound for protocols operating above a critical rate. 

In this paper I show that the results of \cite{renes_duality_2018} and \cite{hayashi_precise_2015} together imply a lower bound on the error exponent for coding over CQ channels which, for suitably symmetric channels, matches the sphere-packing upper bound of Dalai~\cite{dalai_lower_2013}. 
Strangely, then, the current tightest random coding argument for CQ channels in the sense of the error exponent actually comes from analysis of privacy amplification (at least for symmetric channels)! 
The more immediate relation in \cite{renes_duality_2018} is between privacy amplification and compression of classical data relative to quantum side information, which gives a lower bound on the error exponent of compression which matches the sphere-packing upper bound found by Cheng et al.~\cite{cheng_non-asymptotic_2021}.  
Furthermore, their sphere-packing bound can be translated into an upper bound on the exponential decay of the security parameter for privacy amplification based on linear extractors, which tightens the results of \cite{li_tight_2022-1}.
Hence the problem of determining the exponent of the security parameter at very low rates does have a combinatorial nature, as speculated in \cite{li_tight_2022-1}, as it is inherited from the combinatorial nature of the coding error exponent at low rates.

\section{Mathematical setup}
\subsection{Entropies}
To establish these results first requires some preliminary mathematical setup. 
Recall the Umegaki relative entropy of two quantum states $\rho$ and $\sigma$ is given by $D(\rho,\sigma)=\tr[\rho(\log \rho-\log \sigma)]$. 
Here $\log$ denotes the base two logarithm throughout. 
We require two versions of the R\'enyi relative entropy, one by Petz and the other the minimal version in a certain sense (see Tomamichel \cite{tomamichel_quantum_2016-1} for an overview). 
The Petz version of the R\'enyi relative entropy of order $\alpha$ is 
\begin{equation}
\bar D_\alpha(\rho,\sigma)=\tfrac{1}{\alpha-1}\log \,\tr[\rho^\alpha\sigma^{1-\alpha}]\,,
\end{equation}
while the minimal (or ``sandwiched'') version is 
\begin{equation}
\widetilde D_\alpha(\rho,\sigma)=\tfrac{1}{\alpha-1}\log \,\tr[(\sigma^{\tfrac{1-\alpha}{2\alpha}}\rho\sigma^{\tfrac{1-\alpha}{2\alpha}})^\alpha]\,.
\end{equation}
Observe that $\widetilde D_{\nicefrac 12}(\rho,\sigma)=-\log F(\rho,\sigma)^2$, where $F(\rho,\sigma)=\|\rho^{\nicefrac 12}\sigma^{\nicefrac 12}\|_1$ is the fidelity. 
It is known that $\lim_{\alpha\to 1}\widetilde D_\alpha(\rho,\sigma)=D(\rho,\sigma)$ and that $\alpha\mapsto \widetilde D_\alpha(\rho,\sigma)$ is monotonically increasing~\cite{muller-lennert_quantum_2013} (in fact, the same holds for $\bar D_\alpha$). 
Thus, we immediately have the bound
\begin{equation}
\label{eq:fidelityrelentropy}
F(\rho,\sigma)^2\geq 2^{-D(\rho,\sigma)}\,.
\end{equation}

From these two relative entropy quantities we can define two conditional entropies of a bipartite state $\rho_{AB}$ which will be of use to us, as follows:
\begin{align}
\bar H_\alpha^\uparrow(A|B)_\rho &=\max_{\sigma_B}[-\bar D_\alpha(\rho_{AB},\id_A\otimes \sigma_B)]\,,\label{eq:HSibson}\\
\widetilde H_\alpha^\downarrow(A|B)_\rho &=-\widetilde D_\alpha(\rho_{AB},\id_A\otimes \rho_B)\,.
\end{align}
The optimal $\sigma_B^\star$ in \eqref{eq:HSibson} is known from the quantum Sibson identity~\cite{sharma_fundamental_2013}: For all $\alpha\geq 0$, 
\begin{equation}
\label{eq:sibson}
\sigma_B^\star = \frac{(\tr_A[\rho_{AB}^\alpha])^{\nicefrac{1}{\alpha}}}{\tr[(\tr_A[\rho_{AB}^\alpha])^{\nicefrac{1}{\alpha}}]}\,.
\end{equation}

\subsection{Duality}
\label{sec:duality}
The two conditional entropies are dual in the sense that, for any pure state $\rho_{ABC}$~\cite{tomamichel_relating_2014},
\begin{equation}
\bar H_\alpha^\uparrow(A|B)_\rho=-\widetilde H_{1/\alpha}^\downarrow(A|C)_\rho\,.
\end{equation}
The relative entropy is self-dual. 
Entropy duality implies entropic uncertainty relations between conjugate observables~\cite{coles_uncertainty_2012}. 
For a $d$-level quantum system $A$, let $\{\ket{z}\}_{z\in \mathbb Z_d}$ be an arbitrary basis, and define the observables $Z_A=\sum_{z\in \mathbb Z_d} \omega^z \ketbra z$ with $\omega=e^{2\pi i/d}$ and $X_A=\sum_{x\in \mathbb Z_d} \ket{x+1}\bra{x}$, where addition inside the ket is modulo $d$. 
Abusing notation somewhat, we also denote the random variables associated with the measurement outcomes of the observables by $Z_A$ and $X_A$, respectively. 
We denote the elements of eigenbasis of $X_A$ by $\ket{\tilde x}$.
For any quantum state $\rho_{ABC}$ we then have 
\begin{align}
\bar H_\alpha^\uparrow(Z_A|B)_\rho+\widetilde H_{1/\alpha}^\downarrow(X_A|C)_\rho&\geq \log d\,,\\
\bar H_\alpha^\uparrow(X_A|C)_\rho+\widetilde H_{1/\alpha}^\downarrow(Z_A|B)_\rho&\geq \log d\,.
\end{align}
In fact, these inequalities are saturated for certain quantum states, as detailed in \cite{renes_duality_2018}. 
In particular, for pure states of the form 
\begin{equation}
\label{eq:IRfromPAstate}
\ket{\psi}_{AA'BC}=\sum_{z\in \mathbb Z_d}\sqrt{P_Z(z)}\ket{z}_A\ket{z}_{A'}\ket{\varphi(z)}_{BC}
\end{equation} 
with arbitrary probability distribution $P_Z$ and pure states $\ket{\varphi(z)}_{BC}$ we have
\begin{equation}
\label{eq:entropyduality1}
\bar H_\alpha^\uparrow(Z_A|B)_\psi+\widetilde H_{1/\alpha}^\downarrow(X_A|A'C)_\psi= \log d\,.\\
\end{equation}
Similarly, with $\ket{\psi'}_{AA'BC}=\sum_{x\in \mathbb Z_d}\sqrt{P_X(x)}\ket{\tilde x}_A\ket{x}_{A'}\ket{\theta(x)}_{BC}$ we have
\begin{equation}
\label{eq:entropyduality2}
\bar H_\alpha^\uparrow(X_A|C)_{\psi'}+\widetilde H_{1/\alpha}^\downarrow(Z_A|A'B)_{\psi'}= \log d\,.\\
\end{equation}
These equalities also hold for the usual von Neumann conditional entropy. 

These entropy equalities and others defined using similar entropy quantities also extend to $n$-fold copies of the state $\psi$, with observables $X_{A}^n$ and $Z_A^n$, as well as to outputs of linear functions acting on these random variables (see \cite{renes_duality_2018} for more details). 
In particular, suppose that $\check f:\mathbb Z_d^n\to \mathbb \mathbb Z_d^m$ is a surjective linear function and $d$ prime. Then there exists another linear function $\hat f:\mathbb Z_d^n\to \mathbb Z_d^{n-m}$ such that the function $f:z\mapsto \check f(z)\oplus \hat f(z)$ is invertible, where $\oplus$ denotes the direct sum of the vectors. 
Define $\check Z=\check f(Z_A^n)=\sum_{z^n\in \mathbb Z_d^n} \ketbra{\check f(z^n)}$ and similarly $\hat Z=\hat f(Z_A^n)$, which are observables or random variables on $m$ and $n-m$ systems, respectively. 
The action of $\check f$ and $\hat f$ can be extended to the conjugate basis by making use of the unitary representation of the invertible $f$ itself. 
Being linear, $f$ has a matrix representation as $f(z)=Mz$, and so we have 
\begin{equation}
\begin{aligned}
U_f\ket{\tilde x^n}
&=\sum_z \ket{f(z^n)}\braket{z^n|\tilde x^n}=\tfrac{1}{d^{n/2}}\sum_{z^n}\omega^{x^n\cdot z^n}\ket{Mz^n}=\tfrac{1}{d^{n/2}}\sum_{z^n}\omega^{x^n\cdot M^{-1}(z^n)}\ket{z^n}\\
&=\tfrac{1}{d^{n/2}}\sum_{z^n}\omega^{(M^{-1})^Tx^n\cdot z^n}\ket{z^n}=\ket{\widetilde {(M^{-1})^Tx^n}}.
\end{aligned}
\end{equation}
Letting the action $x\mapsto (M^{-1})^Tx$ define the function $g$, the first $m$ outputs define $\check g$ and the latter $n-m$ $\hat g$. Then, in an abuse of notation, set $\hat X=\hat g(X^n)$ and $\check X=\check g(X^n)$. 
For the state $\ket{\psi}_{\hat A\check A A'^nB^nC^n}=(U_f)_A\ket{\psi}_{AA'BC}^{\otimes n}$, a different version of the entropy uncertainty relation is given by \cite[Theorem 3]{renes_duality_2018}:
\begin{equation}
P_{\text{guess}}(\hat Z|B^n\check Z)_\Psi=\max_{\sigma} F(\Psi_{\hat X A'^nC^n},\pi_{\hat X}\otimes \sigma_{A'^nC^n})^2\,.
\label{eq:PguessmaxF}
\end{equation}
Here $P_{\text{guess}}(\hat Z|B^n\check Z)_\Psi$ is the optimal average probability of guessing $\hat Z$ by making a measurement on $B^n$ and using the value of $\check Z$. 

This quantity is relevant to the task of data compression of $Z^n$ relative to $B^n$ by a linear compression function $\check f$, as precisely $\check f(Z^n)$ and $B^n$ will be available at the decompressor.  
Only $\hat Z$ remains to be determined by the decompressor. 
Meanwhile, the quantity on the righthand side is a security parameter in privacy amplification of $X^n$ relative to side information held in $A'^nC^n$, by means of the linear randomness extractor $\hat g$. 
In the language of error correction, if the function $\check f$ is specified by the $m\times n$ check matrix $H$, so that $f(z)=Hz$ and $f$ is the syndrome function associated to the code, then $\hat g$ is specified using an associated $(n-m)\times n$ generator matrix $G$ which satisfies $HG^T=0$ by $\hat g(x)=Gx$. Note that this is the opposite matrix action to the encoding function of the associated linear error-correcting code, which is the map $b\mapsto bG$ for $b\in \mathbb Z_d^{n-m}$; instead, $\hat g$ is the syndrome function of the dual code.

\section{Error exponents}
\subsection{Data compression with quantum side information}
Using \eqref{eq:PguessmaxF} we can convert a lower bound on the error exponent of privacy amplification from \cite{hayashi_precise_2015} into a lower bound on the error exponent for compression with side information. 
The compression task is specified by the CQ state $\psi_{Z_AB}$ from \eqref{eq:IRfromPAstate}, namely $\sum_z P_Z(z)\ketbra{z}_A\otimes \varphi_B(z)$. 
Dual to this, in the sense above, is the privacy amplification problem specified by the state $\psi_{X_AA'C}$, where the goal is to extract randomness from $X_A$ whose value is independent of the quantum states in $A'C$. 
Equation (33) of \cite{hayashi_precise_2015} gives a bound on the security parameter if $\hat g$ is chosen randomly from a 2-universal family of hash functions. One choice of such hash functions, as noted therein, are surjective linear functions based on Toeplitz matrices. 
Specifically, it is shown that for rates $R_{\textsc{pa}}=\log d (n-m)/n< H(X_A|A'C)_\psi$ (measured in bits), the exponent of the relative entropy as security parameter satisfies
\begin{equation}
\lim_{n\to \infty}\tfrac{-1}n \log D(\Psi_{\hat X A'^nC^n},\pi_{\hat X}\otimes \sigma_{A'^nC^n})
\geq \max_{\alpha \in [1,2]} (\alpha-1)(\widetilde H^\downarrow _{\alpha}(X_A|A'C)_\psi-R_{\textsc{pa}})\,.
\end{equation}
To transform this into a bound on the exponent of the error probability in the compression scheme, combine \eqref{eq:PguessmaxF} and \eqref{eq:fidelityrelentropy} to obtain  
\begin{equation}
\begin{aligned}
P_{\text{guess}}(\hat Z|B^n\check Z)_\Psi
&\geq F(\Psi_{\hat X A'^nC^n},\pi_{\hat X}\otimes \Psi_{A'^nC^n})^2\\
&\geq 2^{-D(\Psi_{\hat X A'^nC^n},\pi_{\hat X}\otimes \Psi_{A'^nC^n})}\,.
\end{aligned}
\end{equation}
Then the error probability $P_{\text{err}}(\hat Z|B^n\check Z)_\Psi=1-P_{\text{guess}}(\hat Z|B^n\check Z)_\Psi$ satisfies
\begin{equation}
P_{\text{err}}(\hat Z|B^n\check Z)_\Psi\leq (\ln 2) D(\Psi_{\hat X A'^nC^n},\pi_{\hat X}\otimes \Psi_{A'^nC^n})\,,
\end{equation}
since $1-2^{-x}\leq (\ln 2)x$ for $x\geq 0$. 
Thus by \eqref{eq:entropyduality1} we have
\begin{equation}
\begin{aligned}
\lim_{n\to \infty}\tfrac{-1}n \log P_{\text{err}}(\hat Z|B^n\check Z)_\Psi
&\geq \max_{\alpha \in [1,2]} (\alpha-1)(\log d-\widetilde H^\downarrow _{\alpha}(X_A|A'C)_\psi-R_{\textsc{pa}})\\
&=\max_{\alpha \in [1,2]} (\alpha-1)(\log d-\bar H^\uparrow_{1/\alpha}(Z_A|B)_\psi-R_{\textsc{pa}})\,.
\end{aligned}
\end{equation}
The rate of the compression protocol is simply $R_{\textsc{dc}}=\log d \frac mn=\log d-R_{\textsc{pa}}$, and therefore we obtain
\begin{equation}
\label{eq:errorexpDC}
\lim_{n\to \infty}\tfrac{-1}n \log P_{\text{err}}(\hat Z|B^n\check Z)_\Psi\geq \max_{\alpha \in [\nicefrac12,1]} \tfrac{1-\alpha}{\alpha}(R_{\textsc{dc}}-\bar H^\uparrow_{\alpha}(Z_A|B)_\psi)\,,
\end{equation}
for all rates $R_{\textsc{dc}}$ above $\log d-H(X_A|A'C)_\psi=H(Z_A|B)_\psi$, where this last equality follows from the von Neumann entropy version of \eqref{eq:entropyduality1}. 

Note that the sphere-packing bound of Cheng et al.~\cite[Theorem 2]{cheng_non-asymptotic_2021} has very nearly the same form:
\begin{equation}
\label{eq:spherepackingDC}
\lim_{n\to \infty}\tfrac{-1}n \log P_{\text{err}}(\hat Z|B^n\check Z)_\Psi\leq \sup_{\alpha \in [0,1]} \tfrac{1-\alpha}{\alpha}(R_{\textsc{dc}}-\bar H^\uparrow_{\alpha}(Z_A|B)_\psi)\,,
\end{equation}
again for $R_{\textsc{dc}}>H(Z_A|B)_\psi$. 
Whenever the optimal $\alpha$ in the sphere-packing bound is at least one-half, the two bounds \eqref{eq:errorexpDC} and \eqref{eq:spherepackingDC} agree. From the analogous behavior for the bounds on the channel coding error exponent, we may surmise that this occurs for rates below a critical value, as at high enough rates zero-error compression potentially becomes possible. 

\subsection{Channel coding}
For suitably symmetric channels, the above result implies a tight lower bound on the coding error exponent. 
Here we use symmetric to describe channels for which there is a simply transitive (or regular) group action on the channel outputs, a unitary group representation $V(g)$ such that for all inputs $z,z'$ there exists a unique group element $g$ such that $\varphi(z')=V(g)\varphi(z)V(g)^*$. 
At the risk of some ambiguity, let us label the group by $z$ itself via $\varphi(z)=V(z)\varphi(0)V(z)^*$. 
The optimal input distribution in the capacity expression for symmetric channels is the uniform distribution, which follows from concavity of the function $P_Z\mapsto I(Z:B)_\psi$. 
Here $\psi_{ZB}=\sum_z P_Z(z)\ketbra{z}_Z\otimes \varphi_B(z)$. 
Hence the capacity of a symmetric channel is simply $\log d-H(Z|B)_\psi$ with uniform $P_Z$ in $\psi_{ZB}$. 

Irrespective of channel symmetry, any data compression scheme for uniform $Z^n$ relative to side information $B^n$ whose compressed output is size $m$ can be used to construct a channel code of rate $(n-m)/n$ with the same or better probability of error. 
Any particular compressor output $\check Z=\check z$ defines a code $\{z^n:\check f(z^n)=\check z\}$, and the error probability of the compression scheme is the average of the (average) error probability for determining $z^n$ from the state $\varphi_B(z^n)$ for $z^n$ in the associated code. 
These states are the channel outputs, so to construct a coding scheme we can just use the code which has the best error probability to transmit messages and use the decompressor as the decoder. 
The capacity considerations above ensure that the resulting codes will achieve capacity for symmetric channels. 

Hence, adapted to the case of channel coding, \eqref{eq:errorexpDC} implies that for any symmetric channel $W$ and rate $R$ below the capacity of $W$
\begin{equation}
\label{eq:errorexpCC}
E(W,R)=\lim_{n\to \infty}\tfrac{-1}n \log P_{\text{err}}(W)\geq \max_{\alpha \in [\nicefrac12,1]} \tfrac{1-\alpha}{\alpha}(\log d-\bar H^\uparrow_{\alpha}(Z_A|B)_\psi-R)\,,
\end{equation}
where $Z_A$ is uniform. 
To compare with existing results, it is easiest to convert this expression into the form of the reliability function~\cite{burnashev_reliability_1998}:
\begin{align}
E_0(s,P,W)&=-\log \tr[(\sum_z P_Z(z)\varphi_B(z)^{\nicefrac1{1+s}})^{1+s}]\,,\\
E_0(s,W) &= \max_{P} E_0(s,P,W)\,.\label{eq:reliabilityopt}
\end{align}
A somewhat tedious calculation gives, using \eqref{eq:sibson}, 
\begin{equation}
\begin{aligned}
\log d-\bar H_\alpha^\uparrow(Z_A|B)_\psi
&=\log d+\tfrac1{\alpha-1} \log \tr[\psi_{ZB}^\alpha (\id_Z\otimes \sigma_B^\star)^{1-\alpha}]\\
&=\tfrac1{\alpha-1} \log \tr[\psi_{ZB}^\alpha (\pi_Z\otimes \sigma_B^\star)^{1-\alpha}]\\
&=\tfrac\alpha{\alpha-1} \log  \tr[(\sum_z \tfrac1d \varphi_B(z)^\alpha)^{\nicefrac1\alpha}]\,.
\end{aligned}
\end{equation}
Hence $\log d-\bar H_\alpha^\uparrow(Z_A|B)_\psi=\frac{\alpha}{1-\alpha}E_0(\frac{1-\alpha}\alpha,P',W)$, where $P'$ is the uniform distribution. 
Equivalently, $E_0(s,P',W)=s(\log d-\bar H_{\nicefrac1{1+s}}^\uparrow(Z|B)_\psi)$.

As with the capacity, the uniform distribution $P'$ is optimal in \eqref{eq:reliabilityopt} for symmetric $W$. 
Holevo gives the following necessary and sufficient condition for the optimal $P^\star$ in (38) of \cite{holevo_reliability_2000}: For all $z$,
\begin{equation}
\tr[\varphi(z)^{1/(1+s)}\Big(\sum_{z'} P^\star(z')\varphi(z')^{1/(1+s)}\Big)^s]\geq \tr[\Big(\sum_{z'} P^\star(z')\varphi(z')^{1/(1+s)}\Big)^{1+s}]\,,
\end{equation}
with equality when $P^\star(z)>0$. 
Specialized to the case of uniform $P$, this is just 
\begin{equation}
\label{eq:holevocondition}
\tr[\varphi(z)^{\alpha}\theta^s]\geq \tfrac1d\sum_{z'}\tr[\varphi(z')^{\alpha}\,\theta^s]\,,
\end{equation}
with $\theta=\sum_{z'}\varphi(z')^{\alpha}$ and $\alpha=\nicefrac{1}{1+s}$. 
For symmetric channels we have
\begin{equation}
\begin{aligned}
\tr[\varphi(z)^{\alpha}\theta^s]
&=\tr[\varphi(0)^\alpha (V(z)^* \theta V(z))^s]\\
&=\tr[\varphi(0)^\alpha \Big(\sum_{z'}\big(V(z)^* \varphi(z') V(z)\big)^\alpha\Big)^s]
=\tr[\varphi(0)^\alpha\theta^s]\,,
\end{aligned}
\end{equation}
and \eqref{eq:holevocondition} is satisfied. 
Therefore we may rewrite \eqref{eq:errorexpCC} as 
\begin{equation}
\label{eq:errorexpCC2}
E(W,R)\geq \max_{s \in [0,1]} (E_0(s,W)-sR)\,,
\end{equation}
This bound reproduces the bound found by Burnashev and Holevo \cite{burnashev_reliability_1998} for CQ channels with pure state outputs and compares favorably with Dalai's sphere packing upper bound~\cite[Theorem 5]{dalai_lower_2013},
\begin{equation}
E(W,R)\leq \sup_{s\geq 0} (E_0(s,W)-sR)\,.
\end{equation} 
Moreover, the bounds reduce to the known results for classical channels. 
In light of the relation between the extractor function and the encoding function described at the end of Section~\ref{sec:duality}, the surjective Topelitz matrices mentioned in \cite{hayashi_precise_2015} are also sufficient for achieving \eqref{eq:errorexpCC2}. 
The resulting codewords are systematic encodings of the message, along with a kind of convolution of the message, though here the convolution operation potentially involves the entire message, not just a limited portion of it. 

\section{Polynomial prefactors for the privacy amplification lower bound} 
Theorem 2 of Cheng et al.~\cite{cheng_non-asymptotic_2021} establishes not just an upper bound on the error exponent, but a non-asymptotic lower bound on the error probability itself. 
Defining $E_{\textsc{sp}}(R)=\sup_{\alpha\in [0,1]}\tfrac{1-\alpha}\alpha(R-\bar H^\uparrow_\alpha(Z_A|A'B)_{\psi'})$, they show that for large enough $n$
\begin{equation}
\label{eq:spherepackingDC2}
-\frac{1}n \log P_{\text{err}}(\hat Z|B^n\check Z)_{\Psi'}\leq E_{\textsc{sp}}(R_{\textsc{dc}})+\frac12(1+|E_{\textsc{sp}}'(R_{\textsc{dc}})|)\frac{\log n}n+\frac Kn\,,
\end{equation}
for some constant $K$.

This bound can be applied to the setting of privacy amplification, by starting from \eqref{eq:entropyduality2} and $\ket{\psi'}$ instead of \eqref{eq:entropyduality1} and $\ket{\psi}$. 
The privacy amplification problem is specified by the CQ state $\psi'_{X_AC}$, which describes a fully general prior probability $P_X$ for $X$ and corresponding conditional states $\theta_C(x)$. 
The associated data compression problem is specified by the CQ state $\psi'_{Z_AA'B}$. 
Theorem 3 of \cite{renes_duality_2018} also states that 
\begin{equation}
P_{\text{guess}}(\hat Z|A'^nB^n\check Z)_{\Psi'}=\max_{\sigma} F(\Psi'_{\hat X C^n},\pi_{\hat X}\otimes \sigma_{C^n})^2\,.
\label{eq:PguessmaxF2}
\end{equation}
In terms of the purification distance $P(\rho,\sigma)=\sqrt{1-F(\rho,\sigma)^2}$ involving the actual marginal, this gives 
\begin{equation}
P_{\text{err}}(\hat Z|A'^nB^n\check Z)_{\Psi'}\leq P(\Psi'_{\hat X C^n},\pi_{\hat X}\otimes \Psi'_{C^n})^2\,.
\end{equation}
Combining this with \eqref{eq:spherepackingDC2} gives
\begin{equation}
-\frac1n\log P(\Psi'_{\hat X C^n},\pi_{\hat X}\otimes \Psi'_{C^n})\leq  \frac12E_{\textsc{sp}}(R_{\textsc{dc}})+\frac14(1+|E_{\textsc{sp}}'(R_{\textsc{dc}})|)\frac{\log n}n+\frac K{2n}\,.
\end{equation}
We may express $E_{\textsc{SP}}(r)$ in terms of $\psi'_{X_AC}$ using entropy duality as 
\begin{equation}
 E_{\textsc{SP}}(r)
 =\sup_{\alpha\geq 1}(\alpha-1)(r-\log d+\widetilde H^\downarrow_{\alpha}(X_A|C)_{\psi'})
 \end{equation}
 and then define $E_{\textsc{SP-PA}}(r)=E_{\textsc{SP}}(\log d-r)$ so that we have 
\begin{equation}
E_{\textsc{SP-PA}}(R_{\textsc{pa}})=\sup_{\alpha\geq 1}(\alpha-1)(\widetilde H^\downarrow_{\alpha}(X_A|C)_{\psi'}-R_{\textsc{pa}})\,.
\end{equation}
Furthermore, $E'_{\textsc{SP-PA}}(r)=-E'_{\textsc{SP}}(\log d-r)$, and altogether we have (adjusting the constant $K$)
\begin{equation}
\label{eq:tighterpaupper}
-\frac1n\log P(\Psi'_{\hat X C^n},\pi_{\hat X}\otimes \Psi'_{C^n})\leq  \frac12E_{\textsc{sp-pa}}(R_{\textsc{pa}})+\frac14(1+|E_{\textsc{sp-pa}}'(R_{\textsc{pa}})|)\frac{\log n}n+\frac Kn\,.
\end{equation}
The first term in this expression gives the same $n\to\infty$ limit reported in \cite[Theorem 2]{li_tight_2022-1}; the additional terms give polynomial prefactors to the lower bound on the purification distance itself. Note that this bound is valid only for extractors based on linear functions.

\section{Discussion}
The lower bound \eqref{eq:errorexpCC2} for symmetric channels lends credence to the idea that the sphere-packing upper bound can be matched by random coding lower bounds for all CQ channels (at least for high enough rates). 
New coding arguments are needed. 
Even in the case of symmetric channels we are left in the slightly uncomfortable position of relying on privacy amplification arguments to infer the existence of very good codes. 
It would also be interesting to extend the achievability results to the moderate deviation regime as well as to establish the improved sphere-packing upper bound \eqref{eq:tighterpaupper} for privacy amplification by more direct arguments in order to lift the restriction to linear extractors. 

\section*{Acknowledgments}
I thank Marco Tomamichel and Hao-Chung Cheng for useful discussions. 
This work was supported by the Swiss National Science Foundation through the Sinergia grant CRSII5\_186364 and the National Center for Competence in Research for Quantum Science and
Technology (QSIT).

\printbibliography[heading=bibintoc,title={\large References}]

\end{document}